\documentclass[12pt]{article}
\input epsf.sty
\epsfverbosetrue

\begin{document}

\title{ Inner Market as a ``Black Box''}

\author{Ari Belenkiy
\thanks{E-mail address: belenkiy@albert.ph.biu.ac.il}}

\maketitle
\centerline{\bf Physics Department, Bar-Ilan University, Ramat Gan, Israel.}

\par\bigskip
{\bf Abstract.}

Each market has its singular characteristic. Its inner structure is directly 
responsible for the observed distributions of returns though this fact is 
widely overlooked. Big orders lead to doubling the tails. 
The behavior of a market maker with many or few ``friends'' who can reliably 
loan money or stock to him is quite different from the one without. After 
representing the inner market ``case'' we suggest how to analyze its structure.

\par\bigskip
\par\bigskip
\centerline
{\large\bf Introduction}
\par\bigskip
Recently several market models claimed to find self-organization in market 
behavior. Bak et al (1997) made a surprising claim that one singular feature - mimicking majority - can fix the Hurst exponent of the distribution of price 
returns. An obvious fact was overlooked - namely that an ordinary market 
player cannot know all the prices offered and asked at the market. 
(He knows only the prices of his market-maker.) The only people who 
possess such knowledge are market-makers. 
\par\bigskip
Moreover, only on the level of market-makers one can answer the criticism 
of this model voiced by Levy and Solomon (1997) that the mimicking leads to 
``correlations between investments of large sets of individual investors of 
equal wealth''. Levy and Solomon claimed that such correlations were not 
observed at the level of the ``big market''. However, this is perfectly 
true at the level of the ``inner market''. It is clear that maker-makers 
try to hold their spread as majority do - not to loose money or clients. 
It would be interesting to repeat an experiment of ref. [1] for smaller 
number of agents.
\par\bigskip
Zhang (1999), on the basis of some empirical evidence, claimed that the price 
grows as a square root of market pressure (difference between number of buyers
 and sellers). However, these buyers and sellers, who were taken into account,
 bought and sold their stock at the Bid-Ask prices, or what is the same, 
bought and sold from market-makers. So we can reverse the picture claiming 
that market-makers somewhat ``soften'' the pressure by allowing price to grow 
slower than - as it might be thought - linearly. We will argue that this is 
because their response is somehow delayed.
\par\bigskip
In application of Minority Game to portraying the market dynamics, Giardina et al (2001) argue that random signals (orders) convert a competition between two 
(active and passive) strategies in something similar to the returns of random walk - this leads to the growth of the correlation in activity as square root 
of time. Still various features of such a market remained unexplained: 
Minority Game assumes two outcomes and it is unclear how to distinguish 
between bying and selling. 
\footnote{ Remarks, like ``the best strategies tend to deteriorate'', are in 
contradiction to results of ref. [5].}
As a result, only so-called market activity was investigated. Some important 
things, like truncation of density function for survival time of both 
strategies, remained unexplained.  The major fact, however, which was 
overlooked is that ``strategies'' are the property of the only group in the 
market - market-makers. And two strategies discussed in ref (4) - passive and 
active (buying and selling) - are to hold a fixed 
Bid-Ask spread as other market-makers or to change it (lower Ask or raise Bid).
\par\bigskip
\par\bigskip
\centerline
{\large\bf Inner Market influence on the general market behavior}
\par\bigskip
Solomon and Richmond (2001) provided an excellent example of inner market 
influence on the final market phenomena. Introduction of a specialist 
(Osborne's: the man handling mainly big orders) can double the tail of 
the distribution of the returns. Indeed, from 
matching a pair of two big orders one comes to joint distribution:

$$P1(v>x) = P(v > x) P(v>x).$$

\par\bigskip

This leads from a Pareto tail 
$P(v>x) \asymp x^{-\alpha}$ with $\alpha=3/2$ to a Pareto tail 
$P_1(v>x) \asymp x^{-\beta}$ with $\beta=3$, which is observed across 
the markets.
\par\bigskip
This phenomenon might be also accounted for by an observed ``truncation''
 of the tails of Levy distribution across many markets and is worth further 
investigated. It is interesting that at many markets the specialist is removed at times of crises, so a natural suggestion 
would be to check whether ``truncation'' still exists at the time of crisis. 
\par\bigskip
Another example of possible inner market influence on overall price can be 
seen from Langevin equation in the form developed by Bouchaud and Cont (1998).
This purely 
phenomenological eqaution produced some global market features like bubbles, 
behavior at the time of crisis etc. Surely, their phenomenological description
 of the whole market is also relevant for ``inner market'' - market-makers 
are accustomed to fear $(a)$ and greediness $(b)$ no less than the rest of the
 population.  Then price returns $(u)$ satisfy the equation:

$$Du/dt = -c u +a u - b u^2 + K \kappa(t).$$

To reproduce the empirical laws, especially long-ranged correlations, one must
 be careful about coefficient K before the noise term. 
A natural choice is to put $K = u^2$. 
However, this choice alone does not lead to a power-like tail for distribution
of $u$, as it is easy to see applying Focker-Planck equation 
(ref. [8], p. 230). 
To reproduce a power-like tail of distribution of returns one has to change 
the term $(-c u)$, representing market, to $(-c u^3)$. 
The smaller influence of this term on $Du/dt$ for small u's will reflect 
the fact that small 
trades are completely absorbed by market-makers without changes in prices.  
\par\bigskip
Polynomial tail consists of two parts. Stratanovich's term in the 
Focker-Planck equation brings $u^{-1}$ to the density function and 
the ``market'' term $- c u^3$ will bring $u^{-c}$ and 
together they give $u^{-1-c}$. 
As we saw before, a presense of a specialist may double the constant $c$. 
\footnote{  This, however, should be further tuned to adjust it to the lucid 
``geometrical'' interpretation of different market phenomena (like 
bubbles etc) delineated in the ref. [7]}
\par\bigskip
\par\bigskip
\centerline
{\large\bf Market Basics}
\par\bigskip
Many postulates and nuances of the market ``mechanics'' were explained by 
Osborne (1977), whose book was recently brought to light and popularized 
by McCauley (2000). 
Some of the basic market principles are:
\par\bigskip

1. The market should be continuous (market should suggest a spread at any time).
2. Holding inventories and money, a market-maker should not exceed certain limits. 
3. Profit monies do not enlarge inventory but are put aside.
\par\bigskip
Nuances are: each market-maker could have friends who can allow him to relax his own 
bounds on inventories. This can be conveniently portrayed by a ``friendship''
 matrix, whose entries $G_{i,j}$ show ability by market-maker $j$ to help 
market-maker $i$. (One can chose another approach and assume further that each
 market-maker has his buying and selling curve for buying and selling blocks 
of different size - the exponent already may include the 
``friendship'' matrix.) These two curves - for buying and selling - might be 
slightly different: this asymmetry was emphasized by Zhang (1999).
\par\bigskip
\par\bigskip
\centerline
{\large\bf Market Dynamics}
\par\bigskip
The basic answer to the most empirical laws above is the distribution of  
time intervals for market-makers before changing their spread after trading 
a big amount of stock, or in the language of Minority Game, of changing his 
``passive strategy'' to the ``active'' one.  We assume that a market-maker 
does not like to leave his inventory unbalanced overnight and certainly -  
not over weekend. Therefore he is ready to sell his extra stock for a lower 
price or to buy it back for a higher price. How much lower or higher? 
It depends on the closedness of the market-maker's horizon (week-end). 
(The same feature can be observed at any ``liquid'' fruit-vegetable market - 
on Friday afternoon prices go down dramatically).
\par\bigskip
This scheme implies that the price drops when a market-maker is unable to 
recover symmetry in his inventories for some time depending on his 
``patience'' or ``fear''. We have to evaluate the average time he can hold 
his current spread before going to an ``active'' strategy - to lower his 
Ask or raise his Bid - which automatically will change the market price. 
We should assume that his ``patience'' is also restricted (``truncated'') 
by closedness of the end of the day and even more by the end of the week.  
This ``truncation'' alone can explain power-type tails of the ``passive'' 
time intervals distribution. 
\par\bigskip
The ``root square law'' in activity correlations for small time intervals 
found in Giardina et al (2001) was based on stochastic mathematics suggested 
in Godreche and Luck (1999). 
The major conclusion was that the ``root square law'' comes from the very 
fact of randomness of incoming signals and a resulting $x^{-3/2}$ distribution
 of time switches between two strategies. Our goal is discover the same law 
in the framework described above. 
\par\bigskip
\par\bigskip
\centerline
{\large\bf Mathematical model}
\par\bigskip
We assume that each market-maker $j$ has his maximal asymmetry limit, 
denoted $Limit(j)$
\footnote{ These {\it Limits} are distributed according to Pareto law 
$\asymp w^{-3/2}$.} and 
number of friends which can be encoded in ``friendship'' matrix $G_{i,j}$. 
Randomly, at time t a market-maker trades an amount of stock, denoted 
$Order(t)$, which is also taken from distribution $w^-\alpha$ 
(because orders are proportional to general wealth distribution as 
convincingly, albeit differently, argued in ref. [3], [4], [5]). 
Then, if the $Order$ exceeds his $Limit$, he waits for $\Delta t$ time which 
reflects his ``fear'' and his wealth (ability to accept the $Order$) 

$$ e^{-\Delta t |(Order(t_0) - Limit(j) - \Sigma_i G_{i,j}(t))|} < F(j) \ \ \ \ \ \ \ \ \ \ \ \ \ \  (*),$$

where $F(j)$ is his ``patience'' or ``fear'' 
\footnote{``Fears'' can be thought as uniformly distributed in $(0,1)$ accross j's.} 
and term $\Sigma_i G_{i,j}(t)$ shows a current ability of his friends to help 
him. 
Clearly, if the asymmetry exceeds his limits, he is unable to help his friends.
 At the point in time where the inequality (*) no longer holds, and meantime
the reverse (trade back) order did not occur, a market-maker become 
``active'' - he lowers or raises his spread.  
This time interval might be calculated as 

$$ \Delta t = log(1/F(j)) / |Order(t_0) - Limit(j) - \Sigma_i G_{i,j}(t)|
+ \eta(t_k),$$

where positively distributed $\eta(t_k)$ represents randomly appearing 
opposite orders which can increase his readiness to hold his old spread or, 
if these orders will be comparable to the first one, even cancel his desire 
to change his spread.  
\par\bigskip
Whether the distribution of time holdings of the spread according to this
strategy has $x^{-3/2}$ tail should be further clarified. However, 
this scheme provides a missing ingredient of ref. [4], namely upper truncation 
of the tail of that distribution. The important factor has to be considered - 
the closedness of the end of the day or the end of the week $\delta t$. 
 With unbalanced inventory a market-maker before the week-end likely
changes his spread to recover the balance.  
\par\bigskip
This ``truncation'' was an important part of the argument given in  ref. [4] 
for the variogram of activity to behave as a square root of time for short 
times. 
\par\bigskip
\par\bigskip
\centerline
{\large\bf Further Research}
\par\bigskip
Our next effort will be to uncover the hidden features of the market: number 
of market-makers, their inventory limits and their friendship matrix. 
The question, of course, 
is how. The simplest argument can be borrowed from Solomon and Richmond [6]: 
the lowest limit (wealth) among market-makers might define exponent of the 
distribution of their inventory limits. So from tail $x^{-\alpha}$ one can 
find the lower limit as proportional to $1/(1 - 1/\alpha)$ and thus - 
the wealth of the poorest market-maker. However, many ``friends'' of the 
poorest market-maker can change the tail of the exponent. 
\par\bigskip
We give one more example. Short-ranged correlations in distributions of 
returns (period from 4 to 15 minutes) can be observed across all markets 
(see, e.g., [13, p. 54]). This period shows how quickly the price is 
``recovered'' on average or, in our terminology, how quickly an {\it average} 
market-maker can trade back an {\it average} amount of stock to recover a 
previous spread. So a period from 4 to 15 minutes may uniquely characterize a 
particular market and exact ``inverse'' methods have to be developed to 
uncover hidden (from public) variables. And before answering about average 
strategy of an average player of a ``big market'' we can answer some 
questions about market inner circle.
\par\bigskip
\par\bigskip
\centerline
{\large\bf Conclusions}
\par\bigskip
We suggested a straightforward approach to attack numerous general laws 
observed in time series of price returns. The market, no matter how big, 
is governed by a small group of people called market-makers, whose Pareto-like
 limits for inventories and - more important - particular behavior (their fear
 of unbalanced inventories) predefine distribution laws of the entire market. 
Their ``strategies'' amount to keeping fixed or to changing their spreads by 
different amounts. In either case, their true strategy is to uphold 
fundamental principles of the market.
\par\bigskip
It means that the so-called ``free'' economy is in fact predefined by 
properties of a small group of people who are ready to do their utmost to 
support its spirit and first principles. In social terminology we call them 
bureaucracy or government. In the terminology of ``market ecology'', together 
with ``predates and prey'' (speculators and producers - see [14]) they are 
true hosts of the forest - foresters.
\par\bigskip
\par\bigskip
\centerline
{\large\bf Acknowledgements}
\par\bigskip
Conversations with Y. Avishai, Sh. Havlin, T. Kalisky are acknowledged. 
I admit that the work of J-P. Bouchaud, alone and with collaborators, 
was of great inspiration to this research. Some other important insights 
were learned from the papers of Y-C. Zhang.
\par\bigskip
\par\bigskip
\centerline
{\large\bf REFERENCES}
\par\bigskip
[1] P. Bak, M. Paczuski and M. Shubik. Price variations in a stock market with many agents. Physica A 246 (1997), 430.

[2]. M. Levy and S. Solomon. New evidence for the power-law distribution of wealth. Physica A 242 (1997), 90.

[3] Zhang Y-C. (1999). Toward a theory of marginally efficient markets. Physica A 269 (1999), 30.

[4] I. Giardina, J-P. Bouchaud and M. Mezard (2001). Microscopic models for long ranged volatility correlations. Cond-mat/0105076.

[5] D. Challet, Y-C. Zhang. Emergence of cooperation and organization in an evolutionary game. Physics A 246 (1997), 407.

[6] S. Solomon and P. Richmond (2001). Power laws of wealth, market order volumes and market returns. Cond-mat/0102423.

[7] J-P. Bouchaud and R. Cont (1998). European Journal of Physics B 6, 543 (1998).

[8] N.G. Van Kampen. Stochastic Processes in Physics and Chemistry. Amsterdam, Elsevier 
Science Publ. B.V. 1992.

[9] M. F. M. Osborne 1977. The Stockmarket and Finance from a Physicist's Viewpoint. Crossgar Press.

[10] J. L. McCauley (2000). The futility of utility: how market dynamics marginalize Adam Smith. Physica A, 285 (2000), 506.

[11] C. Godreche and J. M. Luck (2000). Statistics of the occupation time of renewal 
processes. Cond-mat/0010428.

[12] J-P. Bouchaud (2000). Power laws in economy and finance: some ideas from physics. 
Cond-mat/0008103. J-P. Bouchaud and M. Mezard. Wealth condensation in a simple model of economy. Physica A 282 (2000), 536.

[13] J-P. Bouchaud and M. Potters. Theory of Financial Risk. Cambridge UK, 2000.

[14] Y-C. Zhang (2001). Why financial markets will remain marginally inefficient. Cond-mat/0105373

\end{document}